\documentclass[12pt]{iopart}
\pdfoutput=1
\usepackage{fullpage}

\usepackage{iopams,graphicx,color}
\usepackage{setstack}
\usepackage[pdftex,breaklinks=true,bookmarks=false,colorlinks=true,citecolor=blue,urlcolor=blue,linkcolor=black]{hyperref}

\begin{document}

\newcommand{\beq}{\begin{equation}}
\newcommand{\eeq}{\end{equation}}
\newcommand{\beqa}{\begin{eqnarray}}
\newcommand{\eeqa}{\end{eqnarray}}
\newcommand{\ben}{\begin{enumerate}}
\newcommand{\een}{\end{enumerate}}
\newcommand{\hs}{\hspace{0.5cm}}
\newcommand{\vs}{\vspace{0.5cm}}
\newcommand{\todo}[1]{{\color{red} #1}}
\newcommand{\steve}[1]{{\color{magenta} #1}}

\title{Entanglement at a Two-Dimensional Quantum Critical Point: 
a T=0 Projector Quantum Monte Carlo Study}

\author{Stephen Inglis$^1$ and
Roger G. Melko$^{1,2}$}
\address{$^1$Department of Physics and Astronomy, University of Waterloo, Ontario, N2L 3G1, Canada}
\address{$^2$Perimeter Institute for Theoretical Physics, Waterloo, Ontario N2L 2Y5, Canada}

\date{\today}

\begin{abstract}
Although the leading-order scaling of entanglement entropy is non-universal at a quantum critical point (QCP), sub-leading scaling can contain universal behaviour.
Such universal quantities are commonly studied in non-interacting field theories, however it typically requires numerical calculation to access them in interacting theories.  In this paper, we  use large-scale $T=0$ quantum Monte Carlo simulations to examine in detail the second R\'enyi entropy of entangled regions at the QCP in the transverse-field Ising model in 2+1 space-time dimensions
-- a fixed point for which there is no exact result for the scaling of entanglement entropy.
We calculate a universal coefficient of a vertex-induced logarithmic scaling for a polygonal entangled subregion,
and compare the result to interacting and non-interacting theories.
We also examine the shape-dependence of the R\'enyi entropy for  finite-size toroidal lattices divided into two entangled cylinders by smooth boundaries.
Remarkably, we find that the dependence on cylinder length follows a shape-dependent function calculated previously by 
Stephan {\it et~al.}~[New J. Phys., 15, 015004, (2013)] at the QCP corresponding to the $2+1$ dimensional quantum Lifshitz free scalar field theory.
The quality of the fit of our data to this scaling function, as well as the apparent cutoff-independent coefficient that results, 
presents tantalizing evidence that this function may reflect universal behaviour across these and other very disparate 
QCPs in $2+1$ dimensional systems.

\end{abstract}

\maketitle

\tableofcontents
\vfill\eject

\section{Introduction}

At a quantum critical point (QCP),
the R\'enyi \cite{A_renyi} entanglement entropies contain the much-celebrated ability to access the central charge of the associated conformal field theory (CFT) in $1+1$ space-time dimensions \cite{Entang_CFT,ubiquitous_c}.  This has provided the mainstay in a healthy dialog between field theory and numerical lattice simulations, where unbiased methods such as density matrix renormalization group (DMRG) \cite{White92} are able to calculate the R\'enyi entropies of order $\alpha$ precisely in quantum models in one spatial dimension ($1D$).  The form for systems with open ($\eta = 2$) or periodic  ($\eta = 1$) boundaries in $1D$ \cite{Log_Vidal,Cardy},	
\begin{equation}
S_{\alpha} = \frac{c}{3\eta}\left({ 1 + \frac{1}{\alpha} }\right) \log \left[{ \frac{\eta L}{\pi a} \sin \frac{\pi x}{L}  }\right]  + \cdots,  \label{logsin}
\end{equation}
can be straightforwardly compared to lattice numerics \cite{Taddia}, where measuring both $L$ and $x$ in terms of the lattice spacing $a$ gives this term access to the universal central charge, $c$.  
The simplicity of the entangled boundary in $D=1$ means that this and other geometrical factors, such as occur when R\'enyi indices $\alpha > 1$ \cite{Parity}, can be compared between field theory and numerics to a high degree of accuracy in $1+1$ space-time dimensions.

A similar success is only beginning to be enjoyed in $D>1$.
Scalable finite-size lattice simulation methods which can address behaviour at critical points (where the length scale diverges), 
such as quantum Monte Carlo (QMC),
predominantly measure the strongest signal from a looming non-universal leading-order contribution proportional to the boundary length \cite{Sorkin,Shredder},  $L^{D-1}/a^{D-1}= \ell$ (except in the presence of a Fermi surface, where even higher entanglement is expected \cite{WolfFermi,GioevKlich}).  The subleading terms that exist must be obtained using subtraction of very large statistically-fluctuating contributions from this ``area law'' (unless they can be calculated separately, as in infinite-lattice linked-cluster expansions \cite{NLCTFIM}).  What is left is a universal piece, which may depend on the shape or topological characteristics of the boundary, but is believed to not depend on the entangled volume.  This geometry-dependence can be exploited to access different universal numbers in strongly-interacting models, but the types of geometries amenable to comparison between lattice-model simulations and continuum theories has so far been limited.  

One promising function to access universal quantities in 2+1 is the logarithmic contribution that comes from vertices in a polygonal-shaped entangled region  \cite{drum,logcorner,Max,lifshitz_log1,lifshitz_log2}.  The universal coefficients, which depend on the vertex angle $\theta$, have been compared in the past to calculations on finite-size lattices where the entangled region is a square, with four independently-contributing corners ($\theta = \pi/2$).  Particularly relevant are recent numerical results on interacting models \cite{NLCTFIM, TFIM_series, Tommaso}, however to date,
the only field theory calculations that have been performed are on non-interacting theories \cite{logcorner}, inhibiting a quantitative check of universality.

However, other shape-dependent contributions, that occur with smooth boundaries but otherwise impart important geometric characteristics of a 2+1 dimensional space-time geometry, are also accessible by a simulation cell cut into two subregions (e.g.~a torus bifurcated into two cylinders), and reveal sub-leading contributions to the R\'enyi entropy.
Speculation exists about whether this shape dependence reveals an underlying universality \cite{konik,JM_RVB}.  Quantum Monte Carlo simulations on interacting QCPs in $2+1$ dimensions are uniquely poised to answer this question.

In this paper we study the subleading shape dependence of the R\'enyi entanglement entropy 
on
$L \times L$ toroidal lattices with spatial dimension $D=2$, using the critical point of the 
transverse field Ising model (TFIM),
\begin{equation}
H = -J\sum_{\langle i,j \rangle} \sigma^z_i \sigma^z_j - h \sum_{i} \sigma^x_i, \label{TFIMham}
\end{equation}
where ${\overrightarrow \sigma}_i$ is a Pauli spin operator,
accessed by a novel projector QMC algorithm that works strictly at $T=0$. 
Simulating a variety of lattice sizes up to $40 \times 40$ reveals 
universal vertex contributions that are close to a non-interacting field theory in $2+1$.
For smooth boundaries bifurcating the torus into two cylinders, subleading contributions show close functional form to Eq.~(\ref{logsin}) when the two cylinder lengths are associated with $x$ and $L-x$.  However, a cutoff- (or $L/a$-) independent coefficient is not obtained on finite lattices.   
Instead, if we look at the well-studied 2+1 dimensional quantum Lifshitz model \cite{drum,lifshitz1,lifshitz2,lifshitz3} (the field-theory associated with the Rokhsar-Kivelson (RK) Hamiltonian \cite{RK}) there is a proposed functional form \cite{JM_RVB} for the universal subleading term to the area law that is in good agreement with our data with a size-independent coefficient.
This allows us to speculate on the universality of the scaling function, and the interpretation of its coefficient as a measurable witness to the universality class.
Such potential demonstrates the importance of obtaining efficient simulation methods for calculation of R\'enyi entropies in strongly-interacting lattice models, and the continuing need for field theory calculations on interacting fixed points in 2+1 dimensions.

\section{Entanglement at strongly-interacting critical points in 2+1 D }

Extensively studied in one spatial dimension, a multi-disciplinary community is beginning to examine the scaling of entanglement entropy in two (spatial) dimensions, thanks to the rapid development of both theory, and numerical methods for calculating entanglement-related quantities in scalable simulations.
With the important exception of many-body systems housing a Fermi surface \cite{WolfFermi,GioevKlich}, the prevailing paradigm for entanglement in ground state wavefunctions is the area (or boundary) law \cite{Sorkin,Shredder},
\begin{equation}
S_{\alpha} = A \ell + \cdots,
\end{equation}
where $A$ is a non-universal constant, $\ell = L^{D-1}/a^{D-1}$ is cutoff-dependent, and the ellipses indicate a combination of different subleading corrections.  Heuristically, the existence of an area law can be related to the finite extent of correlations across the entangling boundary \cite{wolf,ALreview}.

In gapped systems, the subleading term may have contributions from several constants, some universal, some not.  The most important universal subleading correction gives a contribution dependent on the topology of the entangled surface in a 
fractional topological phases, and is called the {\it topological entanglement entropy}:
\begin{equation}
S_{\alpha} = A \ell - \gamma, 
\end{equation}
where e.g.~$\gamma = \log(2)$ for a simple $Z_2$ spin liquid \cite{LW,KP}.

In 2D critical systems, the behaviour of these subleading corrections to the area law become much more rich.  
One may naively suspect that, due to the diverging correlation length, a violation of the area law might be possible.
However, it can be understood through a course-graining picture that the area law is still obeyed.
First, assume that due to scale-invariance each length scale (in a renormalization group (RG) sense) contributes order $\mathcal{O}(1)$ bit of entanglement entropy across the boundary.
One may take scale-invariance to mean that, when rescaling the system by some factor $b$, the number of modes at this new length scale is proportional to the new boundary length $\ell/b$.  Then, using this assumption and summing over all length scales, an entropy proportional to the boundary length is obtained. \footnote{We thank M. Hastings for pointing out this argument, which might possibly be traced back to J. Preskill.}

At criticality, additional universal subleading terms to this area law are also possible, however they may have a complicated dependence on the geometry of the bipartition.  
Although typically believed, it is not generally known if particular geometric features, for example the number of vertices or the Euler characteristic \cite{drum,cardy-peschel,lifshitz_log1,lifshitz_log2}, give rise to certain universal numbers that can be compared reliably between field theories and quantum lattice models.  
This would be important in making progress towards developing an analog of $c$-theorems \cite{Zamo} in 2+1 space-time dimensions, 
which aim to identify a universal function with monotonic behaviour under RG flows \cite{Tarun}.  
The fact that most field-theoretic calculations are limited to non-interacting systems hampers progress in this regard.
In order to study interacting systems, one must turn to 
numerical techniques on finite-size lattice.  Understandably, the geometries amenable to study on finite-sizes lattice are sometimes different than those that can be studied with continuum field theories, as we now discuss. 

\subsection{Bipartitions with smooth boundaries}

In the continuum thermodynamic limit, dividing a systems into two partitions $A$ and $B$ is easily done with a smooth curved boundary.
This geometry has become particularly important in entanglement monotonicity studies, where
for example, in Lorentz-invariant theories, Casini and Huerta have shown that the entanglement of a smooth circle of circumference $\ell$ scales as $S_1 \sim A \ell - \gamma'$, where the universal constant decreases along an RG flow \cite{circleEnt}.  It is thus a compelling candidate for the monotonic $c$-function.  This may also be related to the entanglement of a three-sphere in odd space-time dimensions, which contains a subleading constant term which changes monotonically along RG flows in holographic theories \cite{Myers1,Myers2}.
Such developments may provide a route to a (2+1)-dimensional analog to the $c$-theorem, especially if the fixed-point value of the monotonic quantity could be determined.

In essentially any interacting theory (and some non-interacting theories) in 2+1, this task would necessarily fall to numerical simulations.  
Unfortunately, such curved geometries are inaccessible on lattices, where obtaining boundaries with sharp vertices is unavoidable when attempting to draw smooth curves.
In the case where the vertices disappear in the continuum, 
it is unknown how such ``pixelization'' might affect the approach to the thermodynamic limit.
In the case where vertices or corners remain in the geometry in the continuum,
they will contribute an additional universal factor, as we discuss in Section \ref{SquareSec}.

Additionally, one may also examine the entanglement entropy across a smooth boundary perturbed away from the critical point, where the correlation length becomes finite.  In this case, even if the boundary has no curvature, one expects \cite{Max},
\begin{equation}
S_{\alpha} = A \ell + r_{\alpha} \frac{L^{D-1} }{\xi^{D-1}},
\end{equation}
for spatial dimensions $D$.  Since the definition of flat boundaries is possible in lattice systems, recent numerical calculations on interacting lattice models have been able to make important comparisons of $r_{\alpha}$ calculated perturbatively in interacting field theories.
In these numerical works, series expansion techniques are able to capture entanglement contributions across flat boundaries on infinite lattices by systematically including larger cluster sizes \cite{TFIM_series,NLCTFIM}.  

An important distinction between these numerical series expansion techniques and quantum Monte Carlo (QMC) is the fact that the latter is typically restricted to periodic finite-size systems (i.e.~toroidal lattices of size $L \times L$), meaning that these methods approach the thermodynamic limit in a different way.  A smooth spatial boundary in a two-dimensional toroidal lattice is possible only if one bifurcates the torus into two separate cylinders.  This geometry can also be studied in certain field theories, which is important for addressing the potential universality of cutoff-independent subleading scaling terms in the R\'enyi entropies.  In this paper will we study this geometry in the context of the TFIM 2+1 QCP in great detail.

\subsection{Bifurcated torus: two-cylinder entropies} \label{twocylinderTHEORY}

\begin{figure}
\begin{center}
\includegraphics[width=0.5\columnwidth]{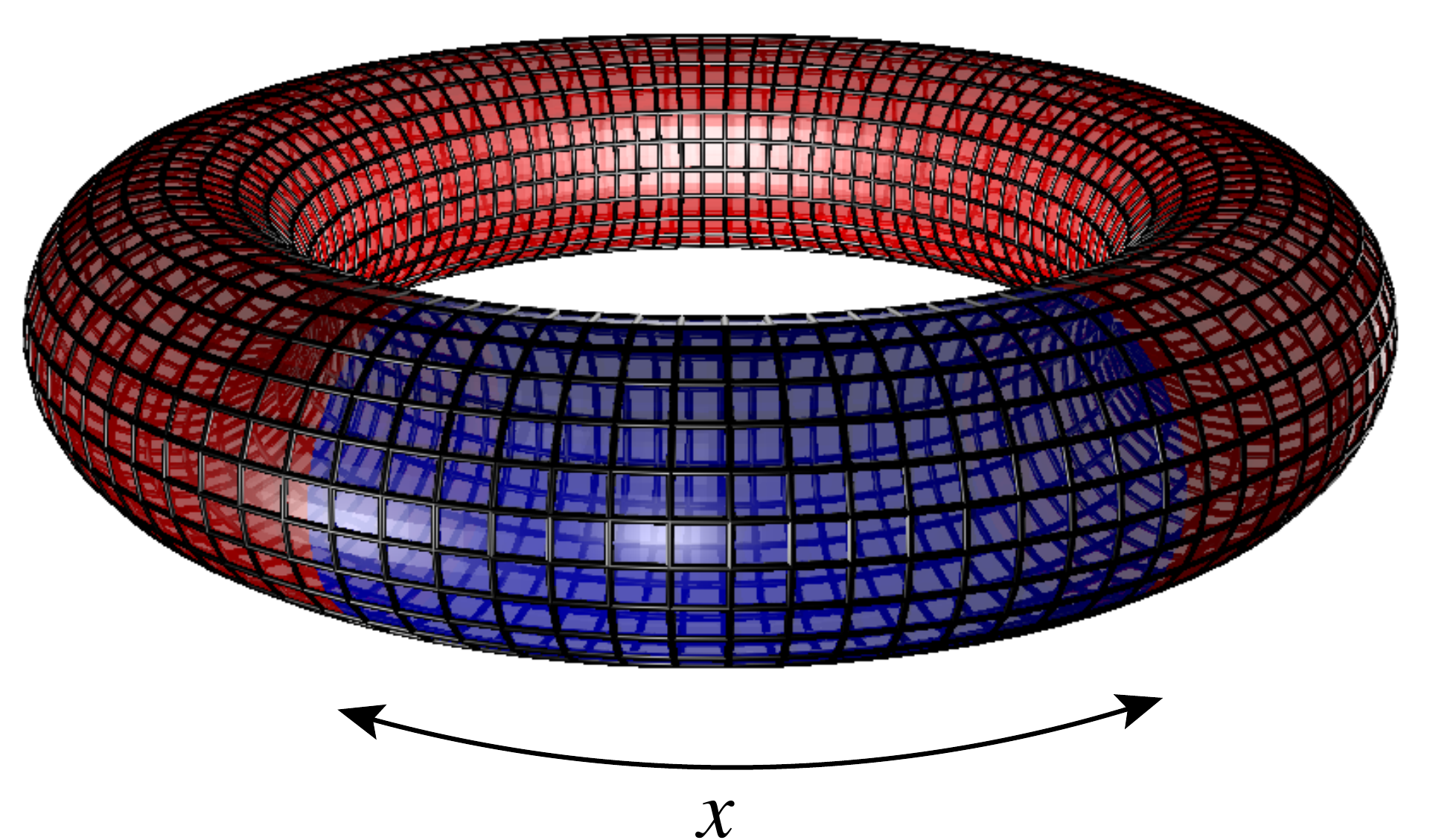}
\caption{A toroidal simulation lattice, divided into two entangled cylinders $A$ and $B$, of size $x \times L,$ and $(L-x) \times L$.
\label{torus1}}
\end{center}
\end{figure}

As discussed above, in two spatial dimensions QMC simulations typically take place on toroidal lattices of size $L \times L$.
For the measurement of entanglement between two subregions $A$ and $B$, a simple geometry is 
illustrated in Fig~\ref{torus1}, where upon bifurcation two cylinders of ``length'' $x$ and $L-x$ are produced.  Past numerical studies of strongly-interacting gapless systems have demonstrated that this geometry is a sensitive probe of the entanglement structure of the wavefunction \cite{KallinHeis,Hyejin,JM_RVB}.  Two possible features are particularly prominent as one varies the length $x$: a striking even-odd effect which arises due to dimer-like physics in the wavefunction; and, a smooth $x$-dependent 
curvature that may contain universal scaling behaviour.
This smooth $x$-dependent curvature is not present in gapped states, making it a sensitive indicator for gapless behaviour in general wavefunctions
\cite{Hyejin} (see Section \ref{CFTsec}).

\subsubsection{Even-odd effect:} \label{evenodd}
As first observed in Ref.~\cite{Hyejin}, 
a striking ``even-odd'' branching effect is observed in $S_n(x/L)$ in certain RVB wavefunctions. 
In Ref.~\cite{JM_RVB}, this effect was understood in a free scalar field theory in the 2+1 dimensional quantum Lifshitz model, which is the field-theory of a Rohksar-Kivelson (RK) Hamiltonian.  It arises due to the underlying dimerization of the wavefunction - and not from any underlying symmetry breaking, as one might naively expect for example in a valence-bond solid phase.

Away from this exactly-soluble free field theory, the even-odd effect serves as a sensitive probe of the degree of dimerization in the low-energy effective description of the wavefunction.  For example, the N\'eel ground state of the 2D spin-1/2 Heisenberg model can be described in an RVB singlet-basis where {\it long} singlets, decaying as $1/r^3$, occur \cite{Sandvik}.  Correspondingly, no even-odd branching effect is observed in the R\'enyi entropy $S_2$ \cite{KallinHeis}.  Similarly, for QCPs which are described by theories ``sufficiently'' far from RK-like Hamiltonians, it's reasonable to expect that no even-odd branching occurs.  As we will see in this paper, in the 2+1 dimensional QCP in the transverse-field Ising model (TFIM), no such even-odd branching occurs.

\subsubsection{Length dependence from conformal field theories:} \label{CFTsec}

Previous numerical studies of interacting models on two-cylinder entropies show
a clear geometry-dependent function, that occurs even in the presence of even-odd branching (described above), and which depends on the cylinder length $x$.  Previous results on the N\'eel ground state of the Heisenberg model, and the square-lattice RVB wavefunction showed reasonably good numerical fits \cite{Hyejin} to the shape-dependence motivated from 1+1 CFT in Eq.~(\ref{logsin}), specifically,
\begin{equation}
S_{\alpha} =  A \ell + b \log(\ell)+ c \log(\sin(\pi y)) +  d, \label{1Dloggeneral} \\ 
 \end{equation}
 where however $b \neq c$ in general.  The term proportional to $\log(\sin(\pi y))$ is heuristically included in order to account for a strong 
 shape-dependence observed in these gapless wavefunctions in 2+1, in analogy to the 1+1 exact result.
Importantly, a non-zero logarithmic term $b$ of order unity was first observed in QMC data on the Heisenberg model in Ref.~\cite{KallinHeis}.  This phenomena was subsequently explain by Metlitski and Grover \cite{MaxTarun} as arising from the presence of Goldstone modes and the restoration of symmetry in a finite-volume simulation cell.  
The coefficient of this additive logarithm should be universal, $b = N_G(D-1)/2$, where $N_G$ is the number of Goldstone modes.
Thus, it is only expected to be present in the case of spontaneously broken continuous symmetry.
It has previously been demonstrated {\it not} to exist in an exactly-solvable finite-size model of free spinless fermions on a square lattice, with $\pi$ flux through each plaquette \cite{Hyejin}.
As we will see in Section \ref{twocylinderFITS}, QMC data for the TFIM quantum critical point is also consistent with $b=0$.

Recently, motivated by the study of dimer RVB wavefunctions that in the continuum limit have conformal invariance in 2D space, CFT techniques have been used to derive an alternate scaling form of the shape-dependent piece for Fig.~\ref{torus1} \cite{JM_RVB, EdEdFend}:
\begin{eqnarray}
S_{\alpha} =& A \ell + c J_{\alpha}(y) + \cdots,  \\
J_{\alpha}(y) =& \frac{\alpha}{1-\alpha} \log \biggl[ \frac{\eta(\tau)^2}{\theta_3(2\tau)\theta_3(\tau/2)} 
 \biggl. \frac{\theta_3(2y\tau)\theta_3(2(1-y)\tau)}{\eta(2y\tau)\eta(2(1-y)\tau)} \biggr], \label{JMtheta}
\end{eqnarray}
where $y = x / L$ is the fractional width of the strip (referred to as $\ell_y / L+y$ in some of the previous literature), $\eta$ is the Dedekind eta function and $\theta_\nu$ is the Jacobi-Theta function.
The above form for $J_{\alpha}(y)$ applies for the R\'enyi entropies with $\alpha \geq 2$, but the definition does not extend to the von Neumann entropy.
Through the rest of the paper we use $J(y) = J_2(y)$ for simplicity.
Due to our geometry (always using $L \times L$ systems), $\tau = iL_x / L_y = i$ never changes in the above equation.
Although this universal function was derived for the quantum Lifshitz field theory, it is interesting to test its universality away from the
critical points describing RK and RVB-like wavefunctions.
In the present paper, we explore its potential for universality by comparing the functional dependence of the R\'enyi entanglement at the TFIM QCP to fits of Eq.~(\ref{1Dloggeneral}) and Eq.~(\ref{JMtheta}).

\subsection{Polygons on a torus: entanglement due to vertices} \label{SquareSec}

\begin{figure}
\begin{center}
\includegraphics[width=0.5\columnwidth]{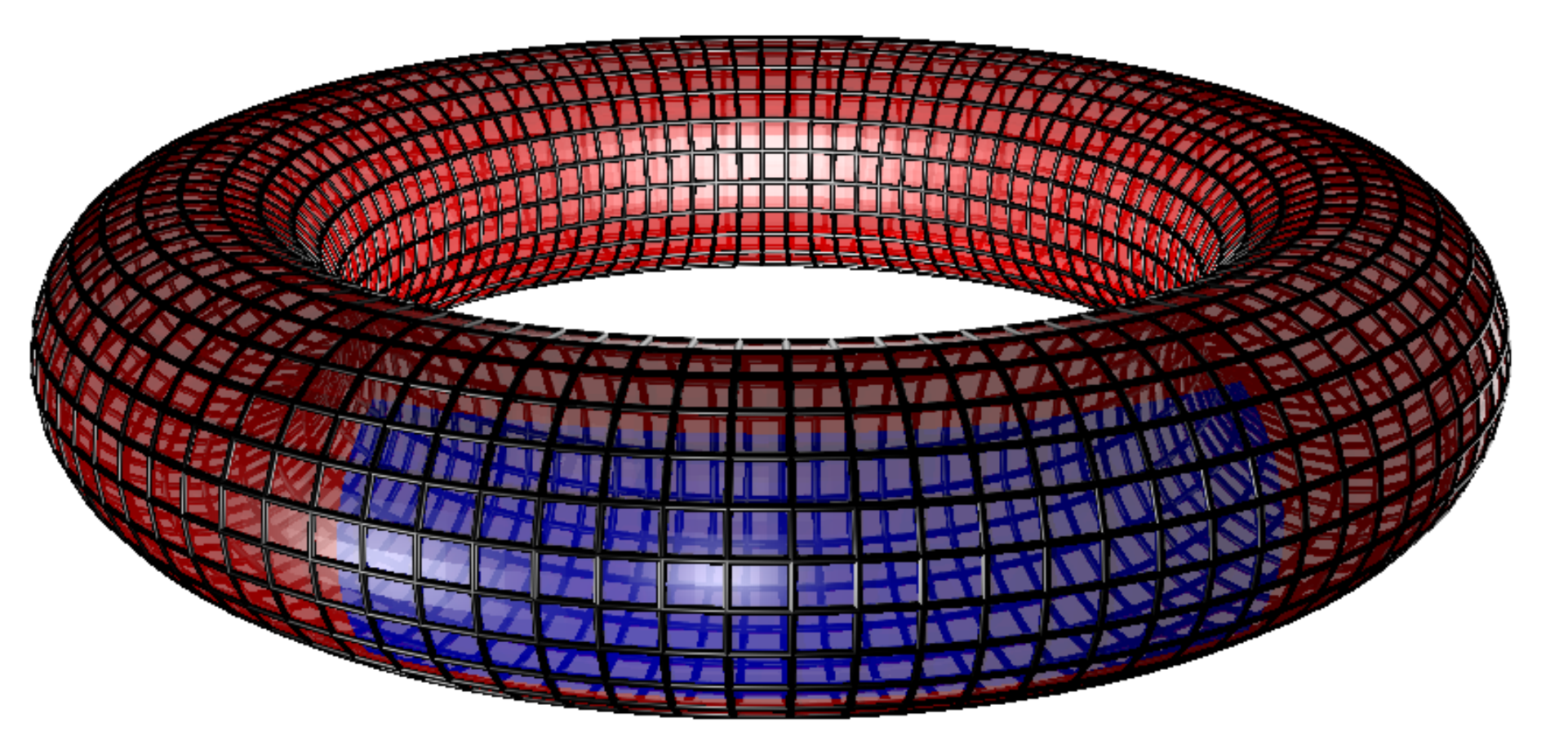}
\caption{A toroidal simulation lattice with a rectangular entangled region A with four $90^\circ$ vertices.
\label{torus2}}
\end{center}
\end{figure}

It is well-known that, in critical systems in 2+1 dimensions, another cut-off dependent contribution to the entanglement entropy
distinct from the boundary (``area''-law) is induced by the presence of vertices or corners.
This term can be seen to have a logarithmic dependence on $\ell$ through heuristic mode-counting arguments in a renormalization group framework.  To begin, one assumes that each length scale (in an RG sense) contributes $\mathcal{O}(1)$ bit of entanglement entropy for each vertex or corner.  Then, summing the contributions from each length scale, one discovers that the corners contribute a constant times the total number of length scales, $\log(\ell)$. \footnote{See footnote 1.}

Refinements and generalization of this argument to other geometrical boundaries exist in the literature.
Most promising for comparison between analytical field theory and lattice numerics is the simple 90-degree vertex.
In Ref.~\cite{logcorner}, the full angle-dependence of the logarithmic vertex contribution is explored, and found to obey,
\begin{equation}
S_{\alpha} = A \ell + n_c a_{\alpha}(\theta) \log(\ell) \label{logfit} + \cdots,
\end{equation}
where $n_c$ is the number of vertices, and $a_{\alpha}(\theta)$ is a universal quantity independent of lattice cutoff.
For the square lattices studied in this paper, it is natural to use a vertex of $\theta = \pi/2$.
The quantity $a_{\alpha}(\pi/2)$
will be universal between the continuum field theory and the lattice model, provided that the angle in the lattice entangled region 
approaches a $90^\circ$ corner as one increases the number of sites to infinity.

Past numerical studies have calculated $a_{\alpha}(\theta)$ for a variety of R\'enyi indices $\alpha$ at the QCP of the TFIM.  
Of relevance to the present study, the value of $a_2(\pi/2)$ has been calculated several times in the past literature.
A value for the non-interacting fixed point in a scalar field theory was provided by Casini and Huerta, $a_2(\pi/2) = -0.0064$ \cite{casiniComm}.  Series expansion studies of the interacting 2+1 dimensional QCP of the TFIM give $-0.0055(5)$ \cite{TFIM_series}
while Numerical Linked-Cluster Expansion (NLCE) gives $-0.0053$; both results are consistent with each other, and lower than that result of the free field theory.  Finally, previous finite-temperature QMC calculations on a single $36 \times 36$-size lattice report $-0.0075(25)$ \cite{Tommaso}.
In the current paper, we aim to improve on these QMC results by employing a new projector method that converges the R\'enyi
entropy $S_2$ directly at $T=0$, as now described.

\section{Projector Quantum Monte Carlo}

Due to a diverging correlation length, the numerical study of quantum phase transitions in strongly-interacting models requires extraordinary care \cite{ARCMP}.
Every effort must be taken to converge data on lattices of as large a size as possible, such that reliable finite-size extrapolations may take place.
Quantum Monte Carlo (QMC) on sign-problem free models provide an unbiased numerical procedure to systematically approach the thermodynamic limit.  For lattice spin models, such as the transverse-field Ising model (TFIM) studied here, the standard procedure used to access quantum critical behaviour involves using highly-efficient {\it finite}-temperature algorithms, such as continuous world-line \cite{Assaad07} or Stochastic Series Expansion (SSE) \cite{SSE1,SSE}, operating at sufficiently low temperatures.  Then, the temperature $T$ is either decreased until it is converged to its $T \rightarrow 0$ behaviour for each lattice size (and parameter set) of interest; or, the inverse temperature $\beta = 1/T$ is set proportional to the linear lattice size, $\beta \propto z L$ (where $z$ is the dynamical scaling exponent), for each lattice size studied.

Recently, a new flavour of QMC algorithm has emerged that combines the simplicity and efficiency of SSE QMC, with the ability to study ground state 
properties of model Hamiltonian directly at $T=0$ \cite{Sandvik}.  Such ``projector'' methods share many of the features of their $T >0$ counterparts, such as: the direct numerical coding of a $D$-dimensional quantum system to a $D+1$ dimensional classical configuration, which is represented and sampled on a computer; and the existence of the prohibitive ``sign-problem'' for fermonic and frustrated systems.  
Unlike $T > 0$ simulations, these methods do not operate in a $D+1$ simulation cell that is periodic in the temporal direction; rather, they operate with a Hamiltonian on a {\it trial} state, repeatedly, projecting out the ground state, as explained below.  These projector methods have been widely adopted for the study of $T=0$ properties of SU(2) (and ${\rm{SU}}(N)$ \cite{SUN_Beach}) invariant Hamiltonians with singlet ground states, providing a simple and efficient method to converge ground state properties \cite{Sandvik,Beach06,AWSloop}.  
In the next section, we describe a procedure, first developed by Sandvik \cite{Sandvik03,unpub} (and recently generalized \cite{Adiabatic}), for the efficient simulation of the U(1) symmetric TFIM Hamiltonian, using an adapted projector QMC method which employs non-local cluster updates.  In Section \ref{swap_sec}, we describe how this algorithm may be adapted to calculate the R\'enyi entanglement entropies using a straight-forward adaptation of the ``replica'' trick \cite{swap}.  It is this method which allows us to accurately study the finite-size scaling of $S_2$ at the QCP of the TFIM, and to access the universal subleading quantities of interest.

\subsection{Algorithm for the transverse-field Ising model}

In 2005, Sandvik introduced a ground state projector QMC method for SU(2) quantum spins, using the so-called valence-bond basis \cite{Sandvik}. 
At $T=0$, QMC methods are tasked with calculating the operator expectation value,
\begin{equation}
\langle \mathcal{O} \rangle  = \frac{1}{Z} \langle \psi^0 | \mathcal{O} | \psi^0 \rangle. \label{zeroExpet}
\end{equation}
Here, one aims to use some procedure to sample $\psi^0$, the ground state wavefunction of a Hamiltonian,  
where the normalization is $Z =  \langle \psi^0 | \psi^0 \rangle$.
The transition probabilities of the Metropolis algorithm are based on this overlap; the non-orthogonality of the 
valence-bond basis ensures that this is trivially non-zero.  However, 
as we will see below, this trial state is not strictly required to be an overcomplete basis.
In the case of the TFIM, an orthogonal $\sigma^z$ basis can be used, provided that wavefunctions are sampled 
such that this overlap is non-zero, using a non-local {\it cluster} algorithm.

In a projector QMC representation, the ground state wavefunction is estimated by a procedure where a large power
of the Hamiltonian is applied to a {\it trial} state, call it $|\alpha \rangle$.  
To see the projection of the ground state wavefunction, one can write the trial state in terms of energy eigenstates $| \psi^n \rangle$, $n=0,1,2 \ldots$,
%\begin{equation}
$|\alpha \rangle= \sum_n c_n | \psi^n \rangle$, so that 
%\end{equation}
a large power of the Hamiltonian will project out the ground state,
\begin{eqnarray}
(-H)^m |\alpha \rangle &=& c_0|E_0|^m \left[{  | \psi^0 \rangle + \frac{c_1}{c_0} \left({ \frac{E_1}{E_0} }\right)^m| \psi^1 \rangle \cdots  }\right], \\
&\rightarrow& c_0|E_0|^m | \psi^0 \rangle \hspace{2mm} {\rm as} \hspace{2mm} m \rightarrow \infty. \nonumber
\end{eqnarray}
Here, we have assumed that the magnitude of the lowest eigenvalue $|E_0|$ is {\it largest} of all the eigenvalues.  To achieve this, one may be forced to add a sufficiently large negative constant to the overall Hamiltonian (that we have not explicitly included).
Then, from this expression, one can write the normalization of the ground state wavefunction, $Z =  \langle \psi^0 | \psi^0 \rangle$ with two projected states (bra and ket) as,
\begin{equation}
Z = \langle \alpha | (-H)^m (-H)^m | \alpha \rangle = \langle \alpha | (-H)^{2m}  | \alpha \rangle, \label{Melko:normZ2}
\end{equation}
for large $m$.  
In a procedure that will be familiar to any SSE aficionado  \cite{DL}, 
the Hamiltonian is written as a (negative) sum of elementary lattice interactions
\begin{equation}
H=-\sum_t \sum_a H_{t,a}, \label{Hdecomp}
\end{equation}
the indices $t$ and $a$ referring to the operator ``types'' and lattice ``units'' over which the terms will be sampled.  
In order to represent the normalization as a sum of positive-definite weights %, $Z = \sum_x W(x)$, 
we can insert a complete resolution of the identity between each $H_{t_i,a_i}$,
\begin{equation}
Z=\sum_{ \{ \alpha\} } \sum_{S_m} 
\prod_{j=1}^{2m}   \left\langle{\alpha_{l}  \left| H_{t_j,a_j}\right| \alpha_r  }\right\rangle.
 \label{Zproj2b}
\end{equation}
Where this equation has been cast in a form similar to that for finite-$T$ SSE.  Note that, 
the sum over the set $\{ \alpha \}$ and the operator list $S_m=[t_1,a_1],[t_2,a_2],\cdots [t_{2m},a_{2m}]$ must be done with importance sampling. 
As we will see below, an update procedure can be constructed that efficiently samples both the list of operators $H_{t,a}$, and (separately) the left and right basis states.  Thus, for sufficiently large $m$, any trial state $| \alpha \rangle$ can be chosen, which is equivalent to using the equal superposition of all spin states $\sigma^z_i$ \cite{unpub}.  Other choices, such as a variationally optimized state, may also be used \cite{unpub}.

Turning to the Hamiltonian Eq.~(\ref{TFIMham}), a convenient definition of operator types is,
\begin{eqnarray}
H_{-1,a} &=& h(\sigma^+_a + \sigma^-_b), \label{TFIM-1a} \\
H_{0,a} &=& h, \\
H_{1,a} &=& J (\sigma^z_i \sigma^z_j + 1).  \label{TFIM1a}
\end{eqnarray}
Note that the index $a$ has two different meanings: a site or a bond, depending on the operator type.  It is evident that some simple constants have been added to the Hamiltonian (Eq.~(\ref{TFIMham})):  the diagonal operator $H_{0,a}$, and also the $+1$ in Eq.~(\ref{TFIM1a}).  The first results in matrix elements with equal weight for both one-site operators.
Denoting each matrix element in the standard basis: the $\sigma^z_i=+1$ eigenstate is $|\hspace{1mm} \bullet \hspace{1mm} \rangle_i$ and  $\sigma^z_j=-1$ is $|\hspace{1mm} \circ \hspace{1mm} \rangle_j$, the non-zero matrix elements are,
\begin{eqnarray}
\langle \hspace{1mm} \bullet \hspace{1mm}  | H_{-1,a} |\hspace{1mm} \circ \hspace{1mm} \rangle = 
\langle \hspace{1mm} \circ \hspace{1mm}  | H_{-1,a} |\hspace{1mm} \bullet \hspace{1mm} \rangle = h, \label{H-1}  \\
\langle \hspace{1mm}  \bullet \hspace{1mm}  | H_{0,a} | \hspace{1mm} \bullet \hspace{1mm} \rangle = 
\langle \hspace{1mm}  \circ \hspace{1mm}  | H_{0,a} | \hspace{1mm} \circ \hspace{1mm} \rangle = h, \label{H0} \\
\langle \hspace{1mm} \bullet \hspace{1mm}  \bullet \hspace{1mm}  | H_{1,a} | \hspace{1mm} \bullet \hspace{1mm} \bullet \hspace{1mm} \rangle = 
\langle \hspace{1mm} \circ \hspace{1mm}  \circ \hspace{1mm}  | H_{1,a} | \hspace{1mm} \circ \hspace{1mm} \circ \hspace{1mm} \rangle = 2J. \label{H1}
\end{eqnarray}
The $D+1$ dimensional projected simulation cell is built such that $2m$ operators of the type \ref{TFIM-1a} to \ref{TFIM1a}  are sampled between the ``end points'' (i.e. the trial states).    Then, sampling occurs via two separate procedures, as follows.  

First, the {\em diagonal update} where one traverses the list of all $2m$ operators in sequence, e.g.~from $| \alpha_{l} \rangle$ to $|\alpha_r \rangle$.
If an off-diagonal operator $H_{-1,a}$ is encountered, the $\sigma^z$ spin associated with that site is flipped but no operator change is made.  If a diagonal operator is encountered, the Metropolis procedure is:
\begin{enumerate}
\item The present diagonal operator,  $H_{0,a}$ or  $H_{1,a}$, is removed. 
\item A new operator {\em type} is chosen, $t=0$ or $t=1$, corresponding to the insertion of either a diagonal $h$ or a diagonal $J$ operator, i.e.~Eq.(\ref{H0}) or (\ref{H1}).  The transition probability to add $H_{0,a}$ is,
\begin{equation}
P(H_{0,a}) = \frac{h N}{hN + (2J)N_b},
\end{equation}
where $N$ and $N_b$ are the number of sites and bonds in the lattice, respectively.  Note, $P(H_{1,a}) = 1- P(H_{0,a})$. 
\item If $H_{0,a}$ is chosen, a site $a$ is chosen at random, and the operator is placed there.
\item If $H_{1,a}$ is chosen, a random bond $a$ is chosen.  The configurations of the two spins on this bond must be parallel for the matrix element to be nonzero.  If they are not, then the insertion is rejected.   Steps (ii) to (iv) are repeated until a successful insertion is made.
\end{enumerate}
One can see that this diagonal update is necessary in order to change the topology of the operator sequence in the simulation cell.  However, in order to get fully ergodic sampling of the TFIM Hamiltonian operators, one must employ non-local updates in addition to these simple diagonal updates.

\begin{figure}
\includegraphics[width=0.9\columnwidth]{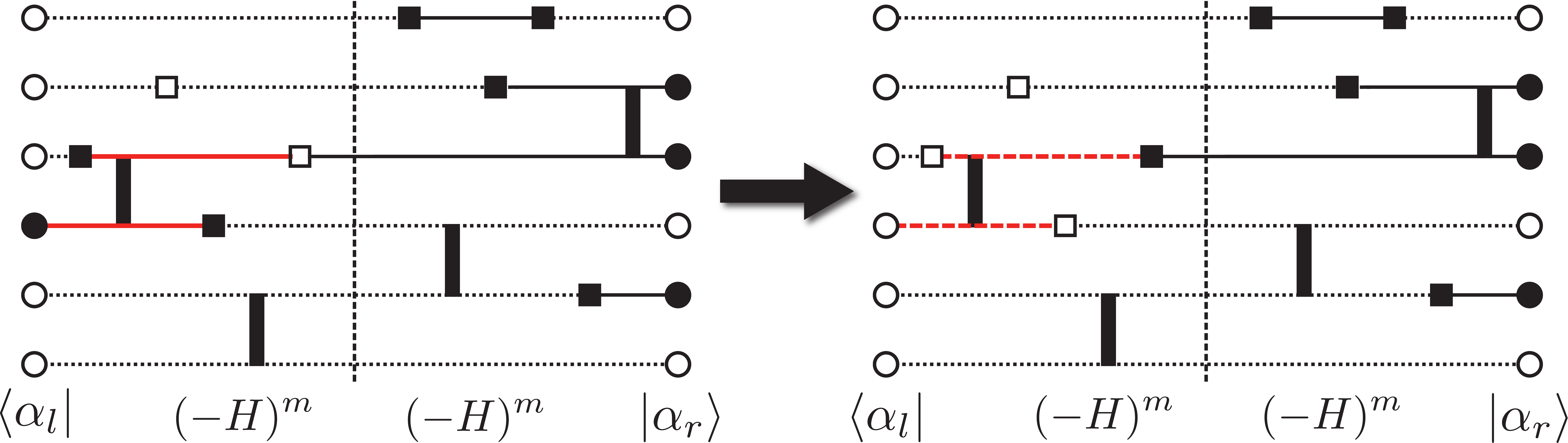}
\caption{Two representations of a 1+1 dimensional simulation cell with $m=6$, before (left) and after (right) a
cluster move that updates site operators and basis states.
The basis states $\sigma^z=1$ ($\sigma^z=-1$) are filled (open) circles, and solid (dashed) lines connecting operators.
Off-diagonal site operators, Eq.~(\ref{H-1}), are represented by filled squares, whereas diagonal site operators, Eq.~(\ref{H0}) are open squares.  Vertical bars represent bond operators, Eq.~(\ref{H1}).
The mid-point of the 1+1 simulation cell, discussed in the text, is indicated by a vertical dashed line.
\label{zeroT_tfim}}
\end{figure}

For $T=0$ projector methods, non-local updates have been discussed previously in the literature in the context of the valence-bond basis \cite{AWSloop}.  
In the present case, the TFIM Hamiltonian does not conserve $\sigma^z$, and a different type of {\it branching} non-local update, called a ``cluster'' update, must be used.  We use cluster updates adapted from the finite-$T$ SSE procedure described in Ref.~\cite{Sandvik03}.
A crucial observation is the judicious choice of $H_{-1,a}$ and $H_{0,a}$ to both have the weight $h$, which allows for unrestricted sampling between the two operator types.  One can easily see that a functional definition of non-local clusters will be an Ising spin forming a space-time group, bounded by either single-site operators, or by spin states of the end point trial states $|\alpha_{l} \rangle$ and $|\alpha_r \rangle$ (see Fig.~\ref{zeroT_tfim}).
If all spins within a cluster are flipped, the total weight of the configuration remains unchanged.  One is then free to build all clusters deterministically, and flip each with a Swendsen-Wang algorithm, i.e.~a probability of 1/2.  In this way, one sees how a fully ergodic sampling of both the operator types and basis state $\sigma^z$ is sampled in the projector QMC.

In Fig.~\ref{zeroT_tfim} special care has been taken to note the mid-point of the simulation projection, since operator expectation values are measured there:
\begin{equation}
\langle \mathcal{O} \rangle = \frac{\langle \psi^0_l | \mathcal{O} |\psi^0_r  \rangle}{Z} = \frac{\langle \alpha_l | (-H)^m \mathcal{O} (-H)^m| \alpha_r \rangle}{ \langle \alpha_l | (-H)^{2m} | \alpha_r \rangle   }. \label{Expect}
\end{equation}
It is particularly important to reiterate, especially in anticipation of the next section, that the wavefunction overlap $\langle \psi^0_l | \psi^0_r \rangle$ must be non-zero (indeed, the Metropolis sampling is set up to force this).  This is ensured by requiring that spin states are the same on the left and the right of the ``mid-point'' of the $D+1$ dimensional cell.  Since spins are only constrained to be the same {\it within} a given cluster, a proper normalization for expectation values could be defined as, 
\begin{equation}
Z=\langle \psi^0_l | \psi^0_r \rangle = 2^{N_0},
\end{equation}
where $N_0$ is the number of independent clusters that cross the boundary.  This follows from the fact that each connected cluster in Fig.~\ref{zeroT_tfim} has two possible spin orientations, $\sigma^z = 1$ and $\sigma^z=-1$, independently.

Various relevant measurements, such as the ground state energy, can be constructed for this simulation \cite{unpub}.  
However, for the purposes of this paper, we are interested specifically in the R\'enyi entanglement entropies, 
which only require knowledge of the cluster structure of the simulation at the middle of the propagated simulation cell; albeit in a 
non-trivial replicated (or multi-sheet) geometry.  Thus, in the next section, we describe the specific implementation of the R\'enyi entropy estimator for the projector QMC for the TFIM.

\subsection{Measuring R\'enyi entropies through the SWAP$_A$ operator} \label{swap_sec}

\begin{figure}
\includegraphics[width=1.0\columnwidth]{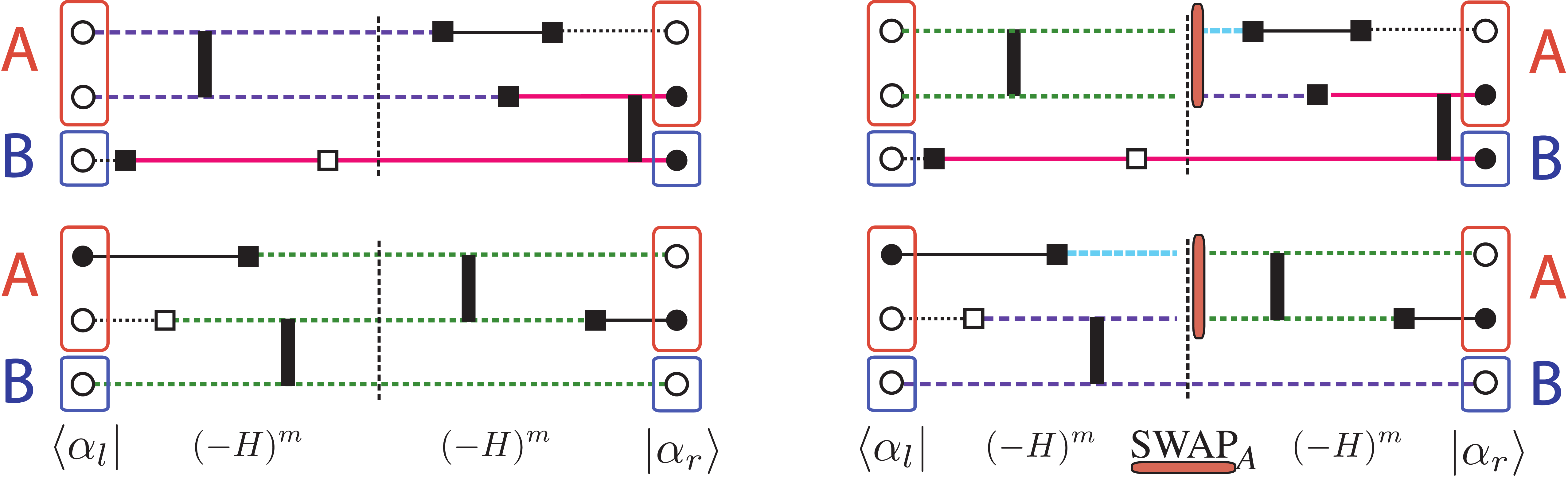}
\caption{An illustration of the ``SWAP$_A$'' operator acting on a replicated simulation cell, for the calculation of $S_2$. 
Basis states and TFIM operators follow the same legend as Fig.~\ref{zeroT_tfim}.  Here, each simulation has three physical spins, and $m=3$ operators in each non-interacting replica, represented one on top of the other.  At left, clusters within each replica that cross the mid-point of the 1+1 simulation cell are highlighted in different colours.  At right, a SWAP$_A$ operator has been applied to permute the basis states corresponding to region $A$, between each replica.  The reconfiguration of the clusters crossing the mid-point can be seen 
from their colours.
\label{swap_tfim}}
\end{figure}

In a development crucial to the study of entanglement at interacting quantum critical points, 
QMC methods have recently been introduced that are capable of measuring the degree of entanglement in a ground state wavefunction, specifically using
the R\'enyi entropies \cite{A_renyi},
\begin{equation}
S_{\alpha} = \frac{1}{1-\alpha} \log \Big[ {\rm Tr}\big( \rho_A^{\alpha} \big) \Big], \label{RenyiEE}
\end{equation}
for integer $\alpha \ge 2$.  Here, $\rho_A$ is the reduced density matrix, $\rho_A = {\rm Tr}_B \{\rho\}$, and $A$ is a subregion of
a lattice system (with $B$ being its complement) -- see Figs.~\ref{torus1} and \ref{torus2}. The von Neumann entanglement entropy corresponds to the limit $\alpha \to 1$.

Unlike a linked-cluster expansion or other methods based on Lanczos diagonalization \cite{NLCTFIM},
the direct measure of R\'enyi entropies can not be done in QMC using conventional estimators.
However, recent work has demonstrated that it is possible to measure
$S_{\alpha}$ for integer $\alpha \ge 2$ using a {\it replica trick} \cite{Cardy,drum,Holz,BP,Naka}.  
For $T=0$, the calculation of $S_{\alpha}$ is done via the following procedure: 
\begin{enumerate}
\item The system is copied into $\alpha$ independent ``replicas''.  
\item Each replica is independently projected to sample the ground state.  The {\it total} wavefunction for the system of all replicas is denoted by $| \psi^0 \rangle$.
\item Following Eq.~(\ref{Expect}),  the operator $\mathcal{O}$ is replaced by the ``SWAP$_A$'' operator for $\alpha=2$ \cite{swap} or permutation operator for
$\alpha \ge 3$ \cite{KallinHeis}.  This operator literally swaps (or permutes) basis states in the region $A$ between the $\alpha$ copies. See Fig.~\ref{swap_tfim}
\item The expectation value 
\begin{equation}
\langle {\rm SWAP}_A \rangle =  \frac{\langle \psi^0_l | {\rm SWAP}_A |\psi^0_r  \rangle}{Z} = 2^{N_A - N_0}, \label{BareSwap}
\end{equation}
where $N_{A}$ is the number of independent clusters crossing the middle of the re-connected (``swapped'') partition function, and $N_0$ is the number of clusters before the swap took place. 
Note: here $N_0$ is the number of clusters in all $\alpha$ replicas or copies.  \label{Step4}
\item Finally, Eq.~(\ref{RenyiEE}) is applied: e.g.~$S_2 = -\log \Big[ \langle {\rm SWAP}_A   \rangle \Big]$
\end{enumerate}

Note that, Step (\ref{Step4}) comes from the fact that Eq.~(\ref{Expect}) is used, with $\mathcal{O} = {\rm SWAP}_A$ as the operator being measured.  This results in the simple procedure of measuring the overlap in the numerator and denominator -- both of which are simply the number of spin states per cluster (2) raised to the power of the number of independent clusters crossing the middle of the $D+1$ dimensional simulation cell.

As first pointed out in Ref.~\cite{swap}, as the lattice size (and particularly the size of region $A$) grows, sampling statistics become exponentially poor when using the naive SWAP$_A$ operator as described above.  Hence, a slight adaptation called the {\it ratio trick} must be used in order to improve statistics.  The ratio trick involves calculating the R\'enyi entropy in several steps, each step being a separate simulation that involves sampling a ratio:
\begin{equation}
\frac{\langle \psi^0_l | {\rm SWAP}_A |\psi^0_r  \rangle}{\langle \psi^0_l | {\rm SWAP}_X |\psi^0_r  \rangle} = 2^{N_A - N_X}  \label{RatioT}
\end{equation}
where $X$ denotes a subregion that is spatially smaller than $A$.  In other words, the weight of a simulation becomes (un-physically) related to a partially-swapped simulation cell.
In a practical simulation, many spatial sub-regions $X_i$, each employed as the weight of a separate simulation, are used to build up towards the physical bi-partition $A$ of interest.  The R\'enyi entropy is then built up by multiplying the contribution of Eq.~(\ref{RatioT}) from each spatial subregion.  

For the TFIM simulation discussed in this paper, the procedure for efficiently calculating the R\'enyi entropy closely follows that used in another $T=0$ projector QMC -- the valence-bond basis QMC for the spin-1/2 Heisenberg model \cite{Sandvik,swap}, which has a detailed description (including the ratio trick) in Ref.~\cite{KallinHeis}.  Remarkably, the only algorithmic difference between the two QMC algorithms is the structure of the space-time clusters employed in the projector method.  In Eqs.~\ref{BareSwap} and \ref{RatioT}, the $N$-numbers ($N_A$, $N_0$, $N_X$) count the number of branching clusters, spanning both the spatial and propagation directions, that cross the middle of the simulation cell.  In the spin-1/2 Heisenberg model \cite{KallinHeis}, this number counts the non-branching {\it loop} structures that cross this middle point, due to the different nature of Hamiltonian operators in that model.  As in the case of the TFIM clusters, each Heisenberg loop in the valence-bond representation has two spin states associated with it (for SU(2); this is modified to $N$ for SU($N$)).  It is remarkable that Hamiltonians with different symmetries, and completely different basis-state representations in the projector QMC, end up with equivalent measurement procedures for the R\'enyi entanglement entropies.

\subsection{Convergence of $S_2$ at a quantum critical point}

\begin{figure}
\includegraphics[width=1.0\columnwidth]{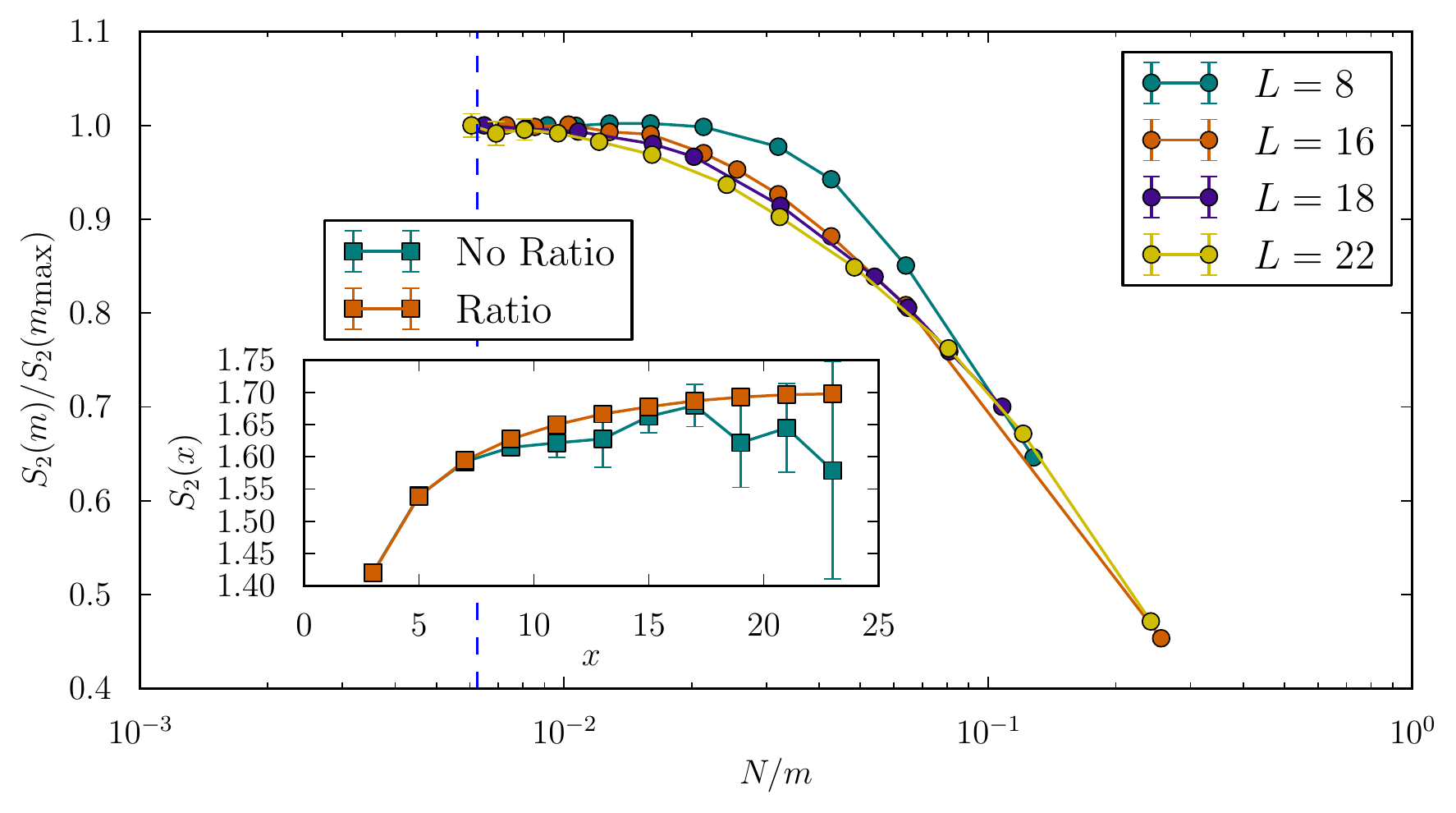}
\caption{The convergence of the second R\'enyi Entropy, $S_2(x = L/2)$, at the centre point of the two-cylinder geometry compared to the value at the maximum $m$ tested, $m_{\mathrm{max}}$, as a function of the number of sites over operators applied to the trial state, $N/m$.
Simulations are performed at the dashed line, $m/N = 160$.
Inset: The second R\'enyi entropy versus the cylinder length, for $N=24 \times 24$.  Note the significant increase in the statistical 
quality of the sampling when the ratio trick, Eq.~(\ref{RatioT}), is employed.
\label{convergence}}
\end{figure}

In this paper, we examine the second R\'enyi entropy, setting $\alpha=2$ in Eq.~(\ref{RenyiEE}).  Like any other observable measured
with the $T=0$ projector method, the value of $S_2$ will have its own unique convergence properties, as a function of the 
projector length $m$, for each value of $h/J$ studied.  When focussing on the critical point, $h/J = 3.044$, where the 
correlation length diverges,  one must be particularly careful to ensure that the simulation is converged in $m$ for each lattice of 
size $N = L \times L$.

In Fig~\ref{convergence}, we examine the value of $S_2$, using the two-cylinder geometry of Fig.~\ref{torus1}, at the point when the two
entangled cylinders are of equal size, $x = L/2$.  Of the geometries studied in this paper, this point is expected to be the most difficult to get good statistics on,
and we use the {\it ratio trick}, Eq.~(\ref{RatioT}) to converge it, building up each subregion $X_i$ with at most $1 \times L$ sites (less for larger sizes).
See also the discussion in the next paragraph for a comparison to data taken without the ratio trick.
For each of the four system sizes that we have studied in detail, we see that the value of $S_2$ at the centre-point $x=L/2$ 
converges for sufficiently large $m$ -- requiring a slightly larger value of $m/N$ as $N$ increases.  For the largest system size that
we collect detailed convergence data on, $L=22$, the value of $S_2$ saturates between $50000<m<60000$ operators, i.e. $m/N$ slightly larger than 100.  We continue to collect data (see Fig~\ref{convergence}) for larger $m$, but find that the value of $S_2$ is essentially converged within 
error bars.  For the data collection in the rest of the paper, we settled on a fixed $m/N = 160$, which is more than sufficient to converge $L=22$ and larger.  A comprehensive study of the $m$-dependence of the R\'enyi entropy for different values of $\alpha$ and $N$ would be an interesting topic
of future study.

In the inset of Fig.~\ref{convergence}, we see raw data for the $x$-dependence of $S_2$ using the geometry in Fig.~\ref{torus1}, on an
$N = 24 \times 24$ lattice with $m/N = 160$.
There, we have compared data obtained from a single simulation using a bare measure of the SWAP$_A$ operator, Eq.~(\ref{BareSwap}), with
that obtained from a procedure where the ratio trick, Eq.~(\ref{RatioT}), is used to build up each entangled region $A$.  Clearly,
naive measurement of the bare SWAP$_A$ operator is insufficient to get controlled statistical sampling for large $x$ values.  We find that, in the 
case of the TFIM at  $h/J = 3.044$, the ratio trick is absolutely necessary for simulation of size $L \ge 16$.

\section{Simulation results on finite-size lattices}

In this section, we report results for the second R\'enyi entropy, $S_2$, for the Hamiltonian Eq.~(\ref{TFIMham}) precisely at the QCP ($h/J = 3.044$) using the QMC simulation method discussed in the last section.  We discuss two entangling geometries: a bipartition between $A$ and $B$ that smoothly cuts each torus into two cylinders (Fig.~\ref{torus1}), and a square bipartition with four 90-degree corners (Fig.~\ref{torus2}).

\begin{figure}
\begin{center}
\includegraphics[width=0.8\columnwidth]{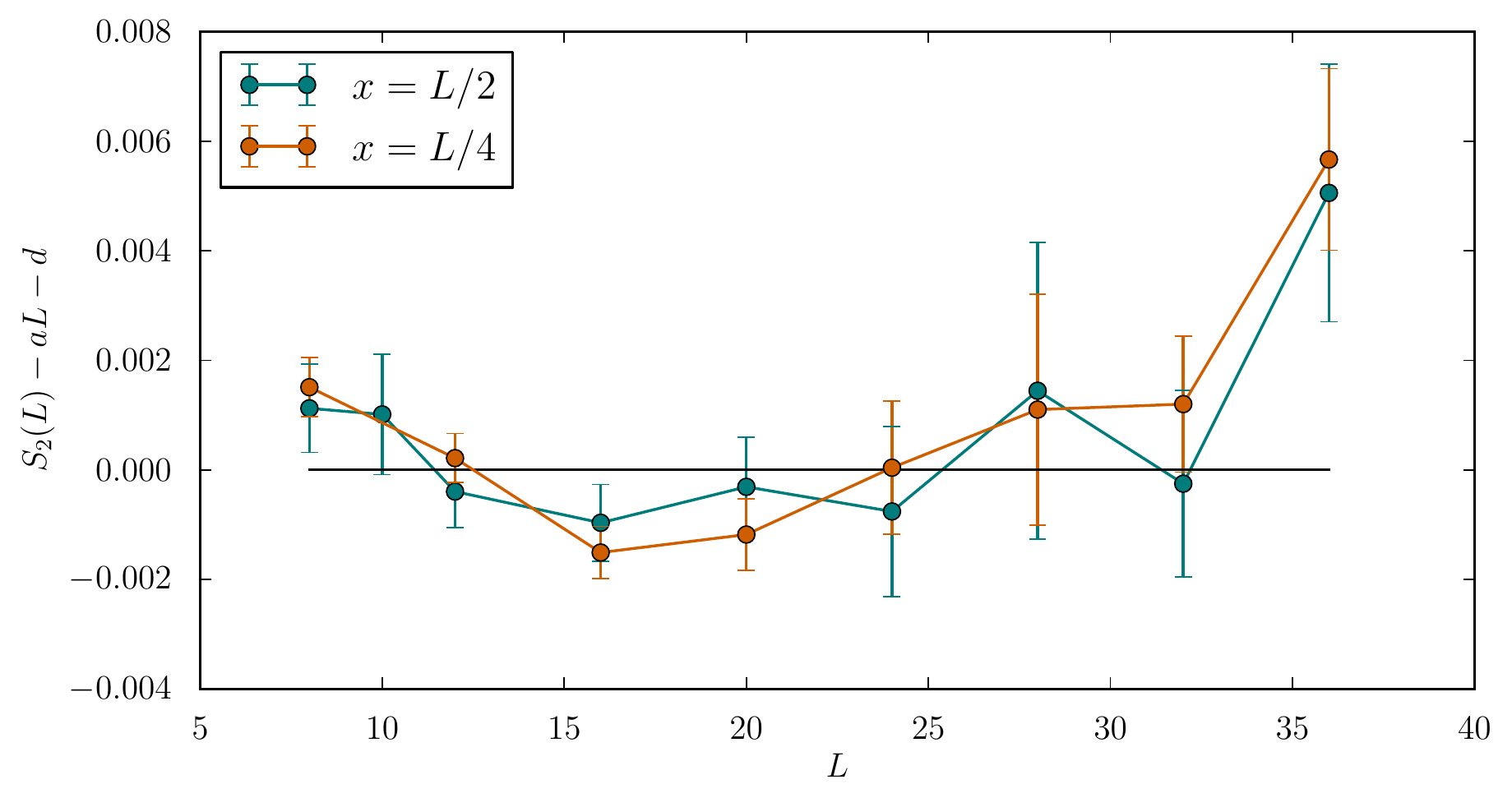}
\caption{(Color online) The residual of fitting all of the $x=L/2$ and $x=L/4$ strips to the form $S_2(L) = aL + d$, keeping in mind $\ell = 2L$ for both geometries.
From this we see that the data fit is consistent with Eq.~(\ref{1Dloggeneral}), with $b=0$ (i.e. no $\log(\ell)$ dependence).
\label{resid_check}}
\end{center}
\end{figure}

\subsection{Bifurcated torus: two-cylinder entropies} \label{twocylinderFITS}

The first geometry we consider is that of an $L \times L$ torus, where the two entangled regions result from smoothly cutting the torus into two cylinders, as shown in Fig.~\ref{torus1}, where the length of each cylinder is $x \times L$ and $(L-x) \times L$.

\begin{figure}
\includegraphics[width=1.0\columnwidth]{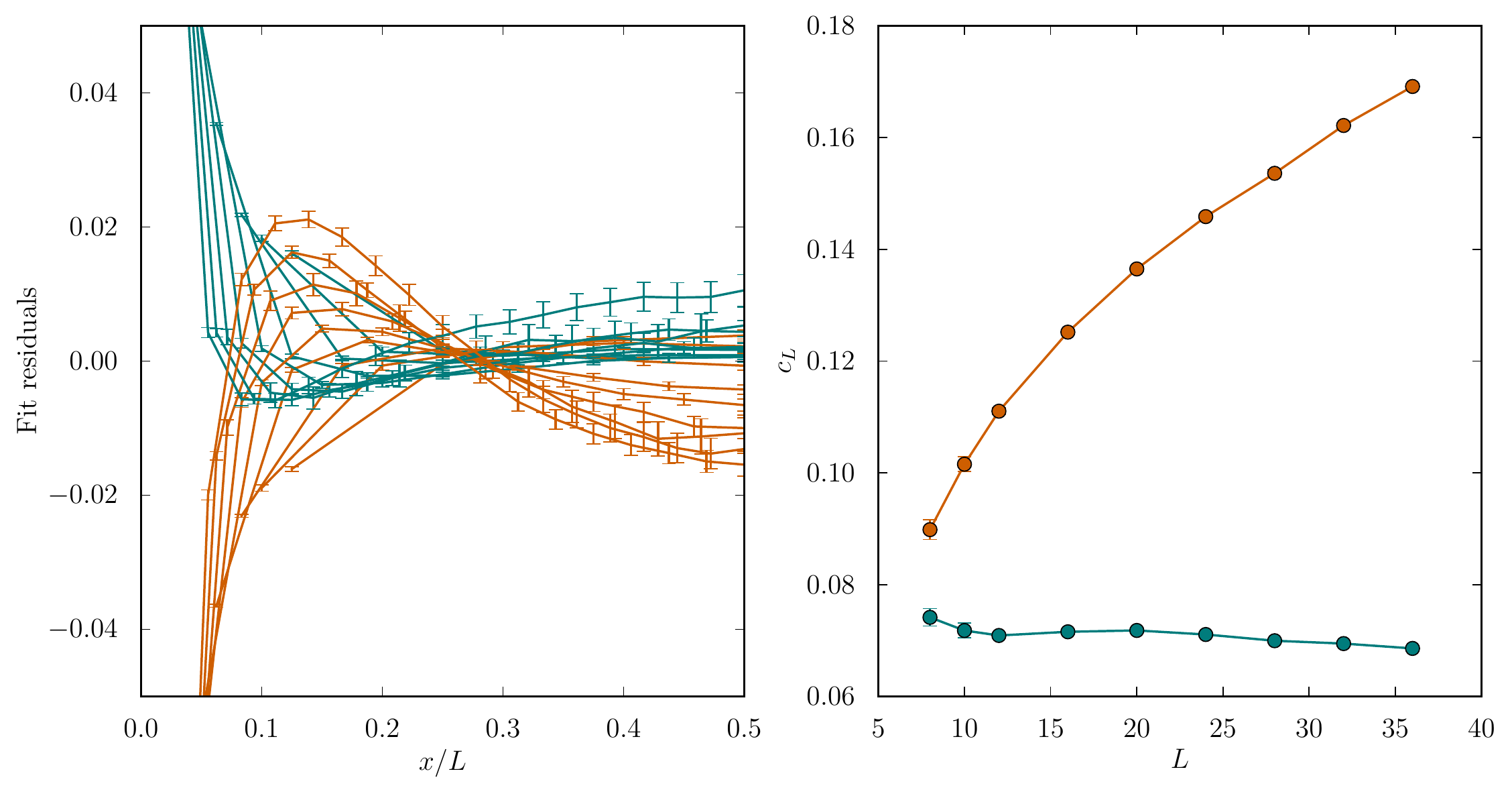}
\caption{ (left) The residual of fitting the entanglement entropy to (orange) Eq.~(\ref{logcompare}) and (teal) Eq.~(\ref{JTcompare}) for all system sizes.
Note the systematic deviation of the redisual in the case of the log fit, that is, the strong tendency to over  estimate (negative residual) the fit at small $x/L$, and under estimate (positive residual) the fit at $x/L$ near $0.5$.
(right) The minimizing parameter $c_L$ for the two fits, with a non-trivial size dependence in the case of the log fit.
\label{fit_comparison}}
\end{figure}

Before examining the full $x/L$ dependence of the two-cylinder geometry, we discuss the possibility of a non-zero $b$ in Eq.~(\ref{1Dloggeneral}) by examining regions $A$ of a fixed $x/L$ embedded in different finite-size lattices $L$.
In Eq.~(\ref{1Dloggeneral}), one may eliminate the $x$-dependence of the term proportional to $c$ by 
fixing $x/L$ to be a constant; e.g. if $x$ is $L/2$ or $L/4$, this contribution will be absorbed into the additive constant $d$.
Fig.~\ref{resid_check} shows the residuals, that is $S_2(L) - aL - d$, for two different choices of $x/L$ as a function of system size for the half and quarter cylinder partitions, with $b$ explicitly set to zero.
From this, we conclude that the fit is acceptable within statistical errors.  
If, instead, we allowed a $b \neq 0$ as a fit parameter, a very small negative coefficient is found, $b \approx -0.01$, 
two orders of magnitude smaller than that found in systems with continuous symmetry breaking \cite{KallinHeis}.
As such, we argue that this value is inconsistent with both physical expectations (see Sec.~\ref{twocylinderTHEORY}) and the results below (see Fig.~\ref{fit_comparison}).
We conclude that our data is consistent with the absence of a  
subleading $\log(\ell)$ dependence in the case of smooth boundaries. 

\begin{figure}
\begin{center}
\includegraphics[width=0.9\columnwidth]{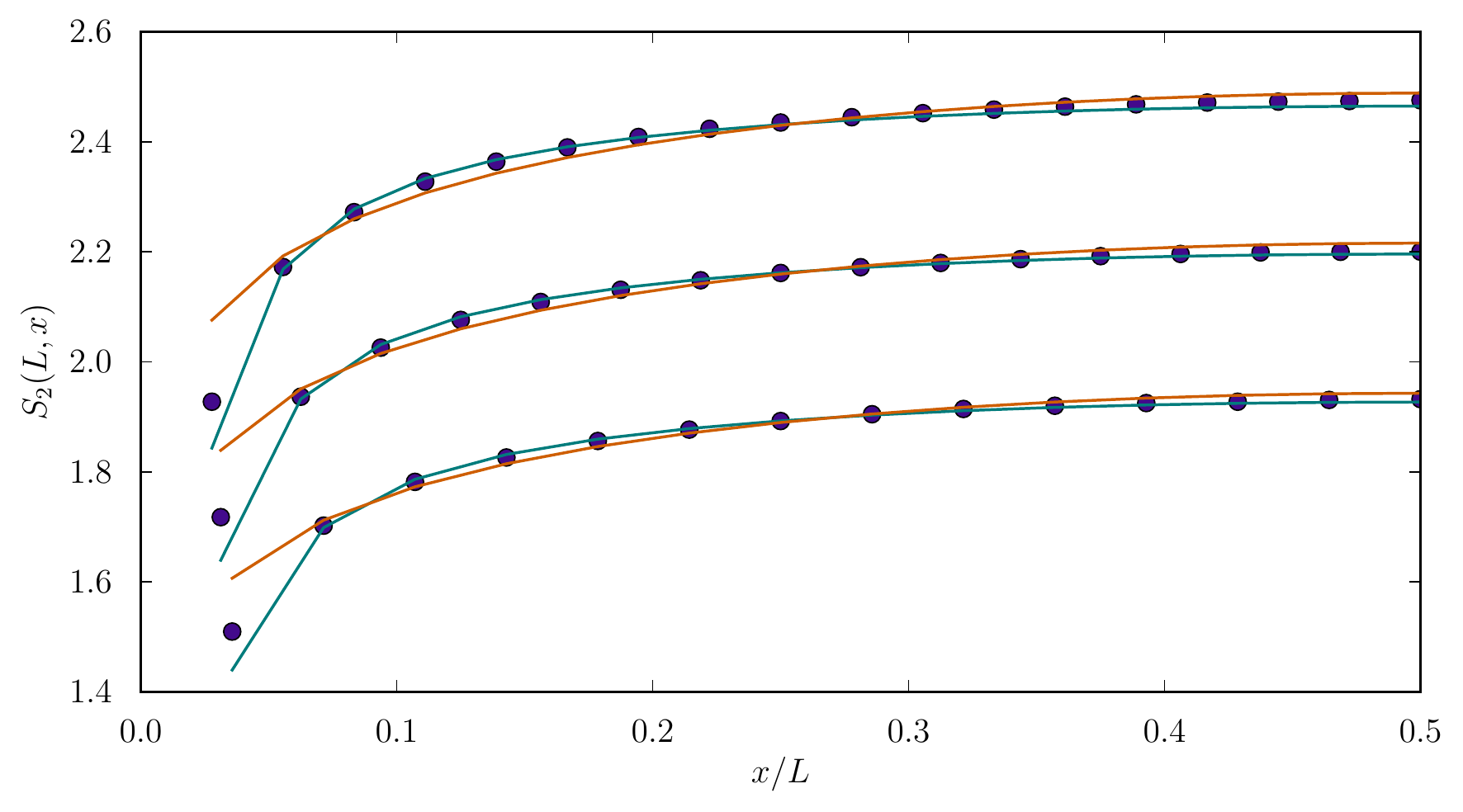}
\caption{(Color online) The entanglement entropy of the $L = 28,32,36$ systems using two cylinders (Fig.~\ref{torus1}), along with fits to (orange) Eq.~(\ref{logcompare}) and (teal) Eq.~(\ref{JTcompare}).
Notice the lack of any even-odd effect in the entanglement as a function of cut length.
\label{no_even_odd}}
\end{center}
\end{figure}

In the following, we assume that no additive logarithms are present in the entanglement scaling at the TFIM quantum critical point; 
$b=0$ in Eq.~(\ref{1Dloggeneral}).
Fig.~\ref{no_even_odd} then shows the entanglement entropy as a function of the cylinder size, $x$, for our three largest system sizes, $L=28, 32$ and $36$.
Recall that in some other gapless wavefunctions, e.g. the RVB wavefunction studied in Refs.~\cite{Hyejin} and \cite{JM_RVB}, a very 
prominent branching effect is apparent which produces separate entanglement curves for even-$x$ and odd-$x$ 
(see Sec.~\ref{evenodd}).
As discussed by Stephan {\it et~al.} \cite{JM_RVB}, this even-odd effect is a measure of how RVB-like a wavefunction is.  In the present case, it is clear that 
if any even-odd effect occurs, it is beyond our ability to detect in these QMC simulations.

Next, we are interested in examining the functional dependence of the shape of the curves in Fig.~\ref{no_even_odd}.
To do this, 
we look at a variety of system sizes, $L$, and a variety of entangled-region lengths, $x$, and examine the fit of the entropy 
to two equations.
The first is the form $\log(\sin(\pi x/L))$, Eq.~(\ref{logsin}), which is relevant for 1+1 dimensional systems, and which has been used
in an ad hoc way in the past analyze some 2+1 dimensional entanglement entropy data \cite{konik}
including systems with a 2+1 dimensional fermi surface \cite{Hyejin} where Eq.~(\ref{logsin}) gives a reasonable (but not perfect) approximation to the fit.
The second is the RVB shape function derived by Stephan {\it et~al.} \cite{JM_RVB}, Eq.~(\ref{JMtheta}).  For the purpose of a numerical comparison, we fit to functions of the form,
\begin{eqnarray}
S_{\log} = & a \ell + c_L \log(\sin(\pi y)) + d, \label{logcompare} \\ 
S_{\mathrm{RVB}} =& a \ell + c_L J(y) + d, \label{JTcompare}
\end{eqnarray}
with $y=x/L$ and where the constants $a$, $c_L$, and $d$ may be different for the above equations.
Note that we allow the freedom that $c_L$ can vary with system size.  However, 
$a$ and $d$ are fit by considering all systems sizes together.
In this way, we can use the variation of $c_L$ with $L$ to examine the quality of fit for each of the two functions.

Fig.~\ref{fit_comparison} shows the residual, $S(L,x) - S_{\log}$ and $S(L,x) - S_{\mathrm{RVB}}$,
of the fit of the entanglement entropy to Eqs.~(\ref{logcompare}) and (\ref{JTcompare}).
In the right panel a comparison of the minimizing $c_L$ for both functional fits is shown.
As is evident from the left-hand plot, the residuals are comparable for the two functional forms, with Eq.~(\ref{JTcompare}) giving a slightly
better fit, using this metric.  However, an important point to notice on the right-hand plot is that, when we attempt to 
fit the data for all system sizes $L$ to the form Eq.~(\ref{logcompare}), there is a clear system-size dependence in $c_L$.  
In Ref.~\cite{konik}, using geometries amenable to DMRG simulation, it was argued that this increasing trend of $c_L$ may level off at 
a moderate system size value; this appears to support the notion of a central charge $c=1$ in the limit of large system size.
As is evident in Fig.~\ref{fit_comparison}, no such conclusion can be drawn from the toroidal QMC geometries for the second R\'enyi entropy -- although, the QMC is not able to probe the behaviour for the von Neumann entropy, making a direct comparison difficult.
In QMC there is no evidence that $c_L$ is converging to a constant value for Eq.~(\ref{logcompare}), leading one to instead conclude that this cannot be a system-size independent (and hence universal) quantity in 2+1 dimensions for the R\'enyi entropy.

In contrast, for the case of Eq.~(\ref{JTcompare}) there is a much less-pronounced size dependence for the coefficient of $J(y)$.  As evident in Fig.~\ref{fit_comparison}, 
a much better case can be made that $c_L$ levels off to a system-size (ie.~cutoff) independent value in the limit of large lattice sizes.
This fuels speculation that 
$J(y)$ may be a universal scaling function relevant for all
fixed points in 2+1 dimensions, as discussed more in Section \ref{discussion}.

\begin{figure}
\begin{center}
\includegraphics[width=0.8\columnwidth]{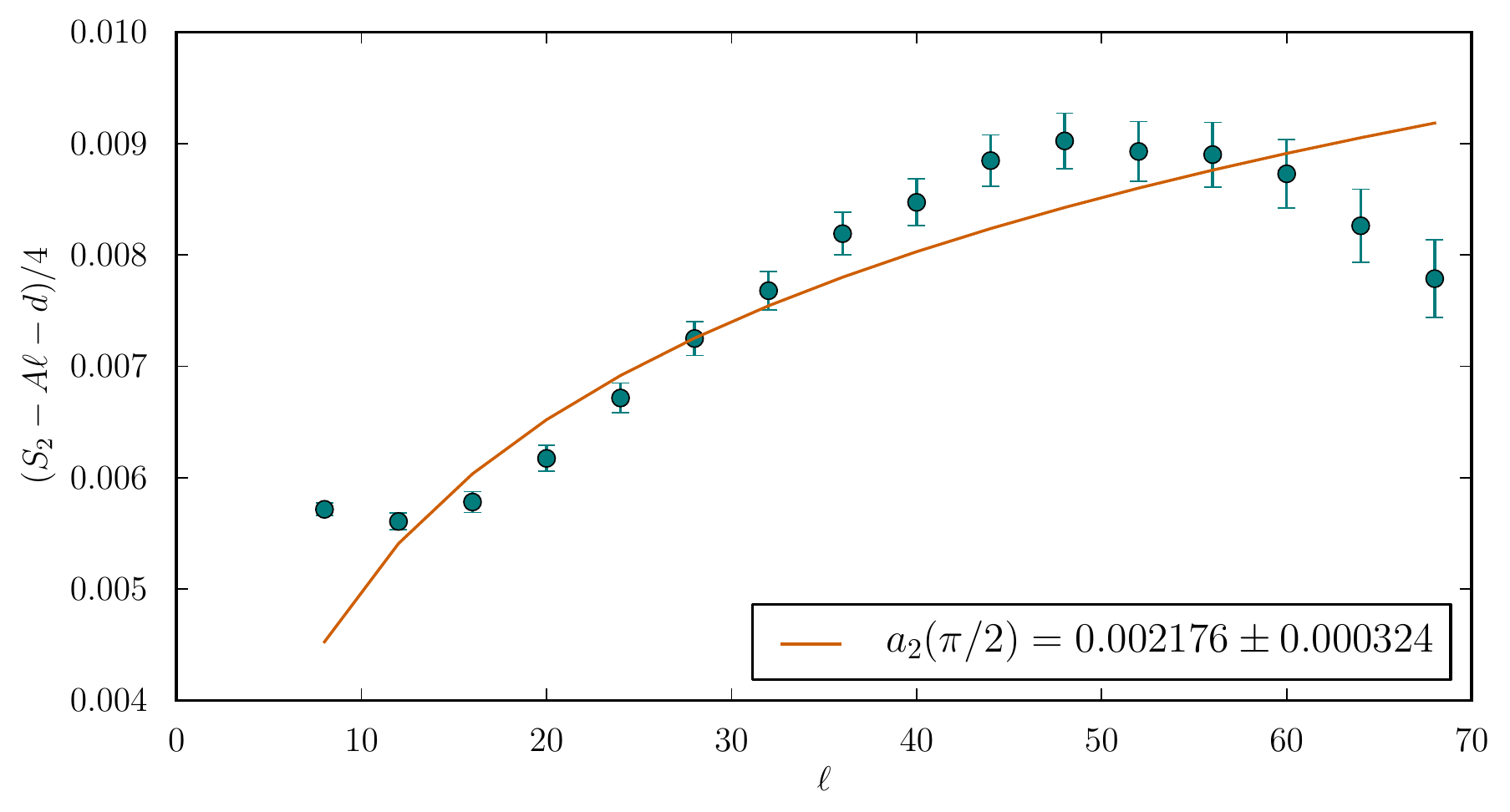}
\caption{The scaling of an $\ell/2 \times \ell/2$ square (of boundary length $\ell$) in a $36 \times 36$ system.
The data is subtracted in such a way that only the log term should remain, with the line showing the best fit to this assumption.
Note that the best fit of the entire data set gives a positive $a_2(\pi/2)$ and there is systematic curvature away from a log, that is to say the deviation of the data seems correlated in $x$ rather than random and gaussian.
$\ell$ here is defined as the (average of inner and outer) boundary length dividing the two regions.
\label{square_compare2}}
\end{center}
\end{figure}

\subsection{Polygons on a torus: entanglement due to vertices}

The second geometry we examine is that of a rectangle embedded in a torus, as shown in Fig.~\ref{torus2}.
As discussed in Section \ref{SquareSec}, this geometry is particularly promising for accessing universal subleading 
corrections to the area law that may be computed in both continuum field theories and lattice models as in this work.
Since these universal corrections arise due to vertices (or corners) in the entangled region, 
any geometry (or aspect ratio) of rectangle should have four separate vertex contributions, 
$4 a_2(\pi/2)$, from Eq.~(\ref{logfit}), assuming that each rectangle is large enough that interactions between corners 
may be neglected.
In our QMC simulation geometries, the use of differently-shaped rectangles gives us 
more than one avenue to extract the coefficient of the subleading log, $a_2(\pi/2)$, thus providing an independent test for universality.

\begin{figure}
\begin{center}
\includegraphics[width=0.8\columnwidth]{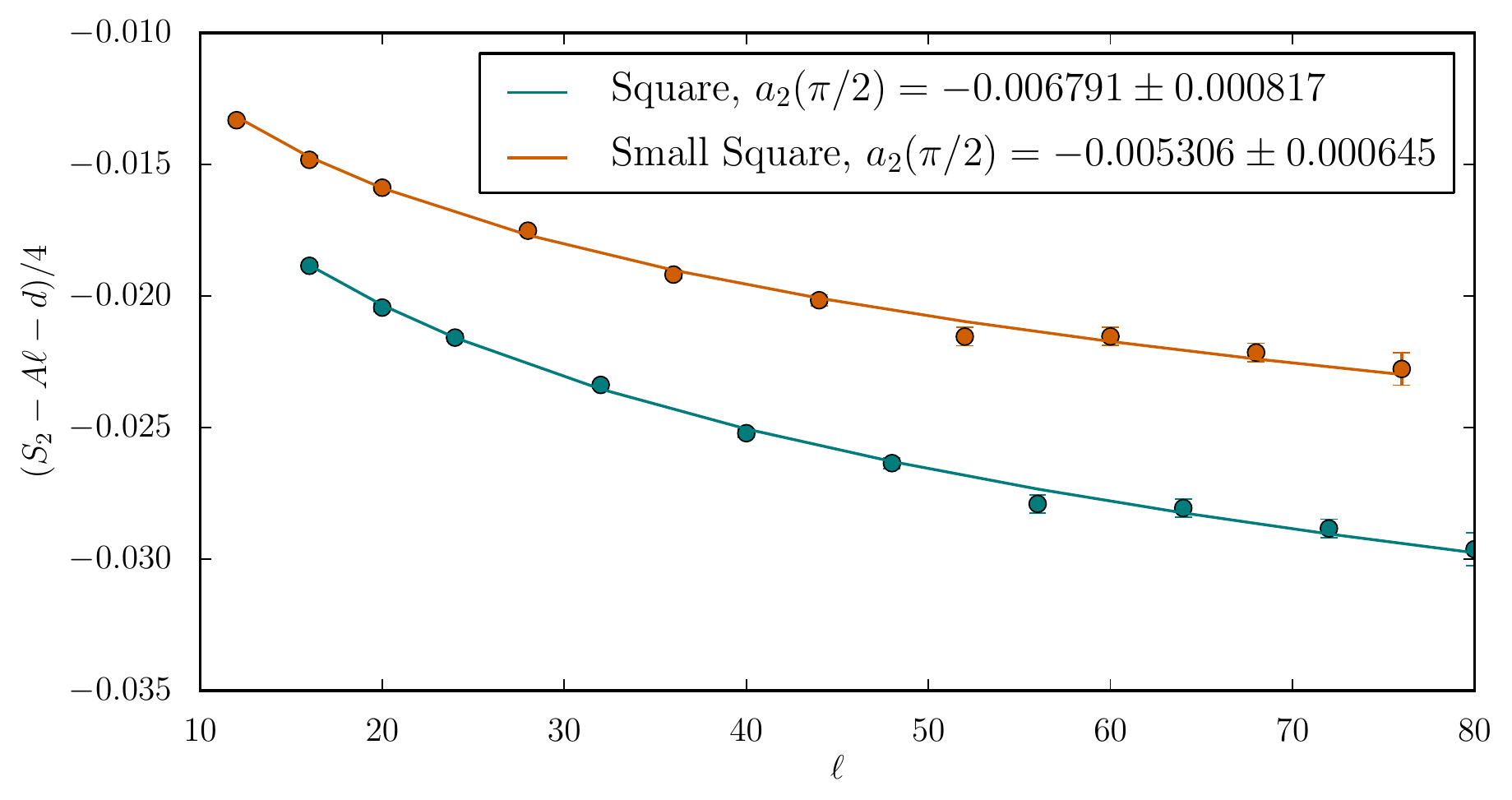}
\caption{
The comparison of scaling an $L/2 \times L/2$ square for two definitions of the boundary length.
``Square'' assumes the boundary is the average of the number of sites on the inside and the outside of the boundary, or simply the length of each edge times four.
``Small Square'' only counts the number of unique sites bordering in the inside edge of the boundary, which is smaller than the previous definition by four sites.
By excluding smaller system sizes the extracted coefficient using these two methods converges, but the error in the extraction also increases.
\label{square_compare}}
\end{center}
\end{figure}

There are two scalings that we test: Fig.~\ref{square_compare2} shows a square scaled in a fixed size simulation, while Fig.~\ref{square_compare} shows an $L/2 \times L/2$ rectangle scaled in an $L \times L$ system.
Scaling of an $L/2 \times L/4$ rectangle was also examined, and results are consistent with the $L/2 \times L/2$ rectangle within error.
We generate the data using the different geometries, then fit it to Eq.~(\ref{logfit}) with an additional allowance for a constant term in the fit:
\begin{equation}
S_2 = A \ell + 4 a_2(\pi/2) \log(\ell) + d.
\end{equation}
Repeating a previous analysis \cite{Tommaso}, we first perform a preliminary fit using a fixed lattice size, $L=36$, and examine the R\'enyi entropy of all possible squares up to size $L/2 \times L/2$.  This data is fit to the form  Eq.~(\ref{logfit}), and the scaling of the subleading logarithmic term as a function of boundary length
is extracted.
The result of this analysis is plotted in Fig.~\ref{square_compare2};
we find that the constant is actually of the opposite sign from that seen in previous work~\cite{Tommaso} (reflecting a significant
decrease in the error bars on $S_2$ with the present data set).
If we exclude smaller rectangles from the fit, the sign of $a_2(\pi/2)$ remains opposite to previous work.
In addition, if we look at the quality of the fit in  Fig.~\ref{square_compare2}, there is a large systematic deviation from a $\log(\ell)$ form, suggesting this geometry may be influenced by other factors not accounted for by this analysis, in particular, terms which depend
on the aspect ratio of region $A$.

In Fig.~\ref{square_compare}, we attempt another type of analysis aimed at eliminating the systematic shape-dependence observed in Fig.~\ref{square_compare2}.  There, the size of region $A$ is fixed to be a square or rectangle with a perimeter proportional to the system's size $L \times L$.  Then, many different $L \times L$ toroidal lattice sizes are individually simulated ($L = 10,12,14,16,20,24,28,32,36,40$).
Fig.~\ref{square_compare}, which shows the subleading logarithm dependence isolated from the area law (and additive constant), demonstrates that this analysis produces a very good fit to the expected 
function $\log (\ell)$.  Analyses of two geometries ($L/2$ and $L/4$) of region $A$ produce a consistent value for the universal coefficient of this logarithm, with the value extracted for squares of size $L/2 \times L/2$ being more accurate (due to the availability of more system sizes).

There are two additional pieces of analysis required for completeness: size exclusion and the definition of $\ell$.
For all results presented thus far on the polygonal entangling geometry, we have taken the definition of $\ell$ to be the average of the number of sites on the inside and outside of the boundary, except where specifically indicated otherwise.
An alternate definition of the boundary length uses the minimum of the number of sites on the inside and outside of the boundary.
These definitions of the boundary length only differ when considering our rectangular regions, differing by four lattice spacings (assuming a square of at least $2 \times 2$ in size).
With very large sets of data, size exclusion can be used to eliminate bias when fitting data for which subleading corrections are not known (here, those proportional to $1/L$).
Since the number of available sizes in our study is limited, we can only do a limited exclusion study.
The result of this analysis is that the corner term tends to become larger as smaller systems are excluded, and with smaller systems included the smaller definition of $\ell$ suggests a smaller (in magnitude) value of $a_2(\pi/2)$.
It should also be noted that with a less comprehensive analysis, it is possible to get values with an artificially lower error bound, and part of the reason for this work is to illuminate the possible pitfalls in the extraction of these universal subleading terms.

All of this analysis suggests a universal corner contribution of $a_2(\pi/2) = -0.006(2)$, with caveats mentioned in the previous paragraph, for the TFIM at criticality, i.e.~the universality class of the Wilson-Fisher fixed point.
Remarkably, this value is very close to the value calculated in a continuum field theory at the {\it non-interacting} Gaussian fixed point by Casini and Huerta, $a_2(\pi/2) = -0.0064$ \cite{logcorner,casiniComm}.  
Previous numerical series expansion studies of the interacting 2+1 dimensional QCP of the TFIM give $-0.0055(5)$ \cite{TFIM_series}
while Numerical Linked-Cluster Expansion (NLCE) gives $-0.0053$ \cite{NLCTFIM}; both results are consistent, and lower than the result from the free field theory.  
Our QMC result gives independent confirmation of the validity of the series and NLCE results; however, due to statistical error bars, it is impossible for the QMC to distinguish between the non-interacting fixed-point value and the previous estimates for the interacting theory.

\section{Discussion} \label{discussion}

Using a novel ``projector'' quantum Monte Carlo (QMC) method that operates at zero temperature \cite{unpub}, we have performed a detailed numerical study of the R\'enyi entanglement entropy, $S_2$, at the quantum critical point of the transverse-field Ising model 
in 2+1 space-time dimensions. 
We have focussed on two different entangling geometries amenable to large-scale simulation on toroidal lattices of size $L \times L$.

First, when the entangling region is a square or rectangular polygon with four 90-degree corners, we confirm the expected scaling form
of
\begin{equation}
S_2 = A \ell + 4 a_2(\pi/2) \log(\ell) + d,
\end{equation}
where $A$ is the non-universal coefficient to the area (or boundary) law, and, 
\begin{equation}
a_2(\pi/2) = -0.006(2),
\end{equation}
is our best estimate of the universal coefficient of the subleading logarithmic term, which arises from each corner.  It is a
very non-trivial success that this value is
within error bars of the value calculated at the same quantum critical point by numerical series and linked-cluster expansions \cite{TFIM_series,NLCTFIM}, both of which take very different approaches to the thermodynamic limit.  
This value of the universal coefficient is also very close to the value calculated in a continuum field theory at the {\it non-interacting} Gaussian fixed point, $a_2(\pi/2) = -0.0064$ \cite{casiniComm}.
The reason for this coincidence may be the fact that the Wilson-Fisher fixed point describing criticality in the TFIM may be reached perturbatively from the Gaussian fixed point.
We hope that this fact motivates future analytical calculations of $a_{\alpha}(\theta)$ in continuum field theory at the interacting Wilson-Fisher fixed point.

Second, we consider the lattice divided into two cylinders of size $x \times L$ and $(L-x) \times L$, separated by two smooth (vertex-less) boundaries  of length $L$.  With this geometry, our data is consistent with the {\it absence} of any ``additive logarithm'', i.e.~no
logarithmic divergence depending on $L/a$.  
This is the same conclusion found in a previous calculation of free spinless fermions on finite-size square lattices, with $\pi$-flux through each plaquette \cite{Hyejin}.
In addition, no even-odd branching effect, like that observed for RVB-like phases \cite{Hyejin,JM_RVB}, is seen in this geometry at the TFIM quantum critical point.
Instead, we find excellent functional fit to
\begin{equation}
S_2 = A \ell + c J(y) + d,
\end{equation}
where $y=x/L$ is the aspect ratio of a cylinder, and 
$J(y)$ (Eq.~\ref{JMtheta})
is a function first derived for the quantum Lifshitz fixed point which describes certain Rokhsar-Kivelson (RK) Hamiltonians.
We argue that this function gives qualitatively better fits to our QMC data, and additionally is consistent with a system-size independent coefficient $c$, which is {\it not} the conclusion one can draw if the data is fit to the familiar form derived for 1+1 CFTs,
$\log(\sin(\pi y))$.  It is interesting to speculate that this new function, $J(y)$, may be a universal scaling function relevant for all
fixed points in 2+1 dimensions, a conjecture that could be addressed with other numerical and field-theoretic studies.  
In particular, QMC simulations should have the ability to calculate the numerical value of the coefficient $c$ at different 
interacting quantum critical points, as demonstrated by the current study.
Finally, we stress the fact that the RVB-shape function $J(y)$ is only applicable for R\'enyi entropies of order $\alpha \ge 2$, i.e.~{\it not} the von Neumann entropy, thus emphasizing the practical utility of $S_2$ in the study of universality at quantum critical points.

We hope that this work, which was obtained with a modest amount of CPU time ($\approx$ 300 CPU-years), will motivate the calculation of similar quantities related to R\'enyi entanglement entropy at a variety of critical points in 2+1 and higher dimensions
in the near future.  In addition, we hope that this demonstration of the most convenient geometries for the study of entanglement in finite-size lattices via quantum Monte Carlo simulations will lead to field-theoretical studies of similar entanglement quantities at a variety of fixed points.  If such a synergy could be accomplished between analytical theory and numerical simulation, the past successes in studying universality through entanglement in 1+1 may soon be
translated to 2+1 and higher-dimensional quantum critical points.

\section{Acknowledgments } We would like to thank A.~B.~Kallin, P.~Fendley, J.-M.~Stephan, E.~Fradkin, R.~Konik, T.~Roscilde, M.~Metlitski and M.~Hastings for enlightening discussions, and N.~Shettell for help with the figures.  We are particularly indebted to A. Sandvik for sharing details of his Projector QMC method, and 
for hospitality during a Quantum Condensed Matter Visitors Program at Boston University.
This work was made possible by the computing facilities of SHARCNET.  Support was provided by NSERC of Canada,
and the Canada Research Chair program (R.G.M., Tier 2).

\section{References}

\bibliography{Biblio}{}

\end{document}